\documentclass[a4paper,11pt]{article}
\usepackage{jheppub} % for details on the use of the package, please see the JINST-author-manual
\usepackage{lineno}
\usepackage{slashed}
\usepackage{amsmath}
\usepackage{xspace} 
\usepackage[]{subfig}
\usepackage[compat=1.1.0]{tikz-feynman}
\newcommand{\ooatau}{\ensuremath{\mathcal{OO}_{1,a_{\tau}}}\xspace}
\newcommand{\ooatauphavg}{\ensuremath{\mathcal{OO}_{1,a_{\tau}}^{\, \lambda_{3},\lambda_{4}}}\xspace}
\newcommand{\ooataufullpol}{\ensuremath{\mathcal{OO}_{1,a_{\tau}}^{\,\lambda_{1},\lambda_{2},\lambda_{3},\lambda_{4}}}\xspace}

\title{\boldmath Probing optimal measurements of the electromagnetic dipole moments of the $\tau$ lepton}

\author{Kartik Bhide$^{1}$,}
\author{Valerie Lang$^{1}$}

\affiliation{$^1$ Albert-Ludwigs-Universität Freiburg, Physikalisches Institut, Hermann-Herder-Straße 3, D-79104 Freiburg, Germany }
 
\emailAdd{kartik.bhide@physik.uni-freiburg.de}

\abstract{Precise measurements of the electromagnetic dipole moments of charged leptons are powerful probes of physics beyond the Standard Model of particle physics. It is essential to study avenues for optimizing the measurement strategies for the anomalous magnetic dipole moment (AMDM) and the electric dipole moment (EDM) of the $\tau$-lepton which have recently become accessible at the Large Hadron Collider (LHC).
In this work, we study the prospects of employing optimal observables for measuring the AMDM of the $\tau$-lepton in PbPb collisions at the LHC. The performance of optimal observables is investigated in a statistical analysis, assuming an integrated luminosity of 2.0~nb$^{-1}$, approximately corresponding to the amount of data collected during PbPb collisions in Run~2 of the LHC.
Improvements of sensitivity to the $\tau$ AMDM with optimal observables are obtained only for a larger phase space than studied so far, more integrated luminosity or including the determination of the helicities of the $\tau$-leptons. Both, chances and limitations, of optimal observables for the $\tau$ AMDM determination will be important to consider in future experiments.}
\begin{document}

% custom commands
\newcommand{\s}{\slashed}
\newcommand{\kbhide}[1]{{\textcolor{red}{{KB:#1}}}}

\maketitle
\flushbottom

\section{Introduction}\label{Sec:Introduction}

High precision measurements of the anomalous magnetic dipole moments (AMDM's) of charged leptons, defined as $a_{\ell}=\left(g-2\right)/2$, have historically been powerful tests of the Standard Model. The AMDM of the electron ($a_{e}$), known to $\mathcal{O}\left(0.1\right)$ ppt precision~\cite{Fan:2022eto}, is one of the most precisely measured predictions of the Standard Model (SM), and is used to define the fine structure constant $\alpha_{EM}$~\cite{Tiesinga:2021myr}. The AMDM of the muon ($a_{\mu}$), known to $\mathcal{O}\left(100\right)$ ppb precision~\cite{Muong-2:2021vma,Muong-2:2023cdq}, has seen a long-standing tension between the SM prediction and experimental measurements, allowing for constraints to be applied on a wide range of Beyond Standard Model (BSM) models (see~\cite{Keshavarzi:2021eqa,Athron:2021iuf} and references therein for comprehensive reviews on the subject).

Compared to the electron and the muon, the AMDM of the $\tau$ lepton has not been measured to a similar precision so far. With a mean lifetime of around $2.9\times 10^{-13}$ s~\cite{ParticleDataGroup:2022pth}, it decays too fast for standard spin precession techniques to be used. Instead, scattering processes at collider experiments where $\gamma\tau\tau$ interactions are present can be used to measure $a_{\tau}$~\cite{delAguila:1991rm,Cornet:1995pw,Atag:2010ja,Galon:2016ngp,Koksal:2018env,Koksal:2018xyi,Gutierrez-Rodriguez:2019umw,Beresford:2019gww,Dyndal:2020yen,Chen:2018cxt}. Other processes involving radiative $\tau$ decays~\cite{Fael:2013ij,Eidelman:2016aih} and bent crystals~\cite{Fomin:2018ybj,Fu:2019utm} have also been proposed previously.

The most stringent bound on $a_{\tau}$ before the LHC-era came from the DELPHI experiment at LEP, where the variation of the total cross section of the $e^{+}e^{-}\rightarrow e^{+}e^{-}\tau^{+}\tau^{-}$ process as a function of $a_{\tau}$ was studied. The limits that were obtained on $a_{\tau}$  at the 95\% confidence level are~\cite{DELPHI:2003nah}:
\begin{equation}
    -0.052 < a_{\tau} < 0.013\,.\label{Eqn:DELPHIlimitsAtau}
\end{equation}
Recently, the AMDM of the $\tau$ lepton has been measured by the ATLAS and CMS experiments, using ultra-peripheral PbPb collisions~\cite{ATLAS:2022ryk,CMS:2022arf} and proton-proton collisions~\cite{CMS:2024skm} at the LHC. Kinematic distributions of $\tau$ decay products were used to extract limits on $a_{\tau}$. The CMS measurement in proton-proton collisions reported the most stringent limits on $a_{\tau}$ to date~\cite{CMS:2024skm}:
\begin{equation}
-0.0042 < a_{\tau} < 0.0062\,.\label{Eqn:CMSlimitsAtau}
\end{equation}
Despite tremendous progress in reducing the limits on $a_{\tau}$ in the LHC-era, current measurements are still far away from the SM prediction~\cite{Keshavarzi:2019abf} of $a_{\tau}=\left(117717.1\pm3.9\right)\times10^{-8}$. 

The electric dipole moments (EDM's) of charged leptons, $d_{\ell}$, can also be used as probes of new physics. Leptons are assumed to be point-like charges, so a non-zero EDM can hint at further substructure of charged leptons. A non-zero EDM can also arise due to higher order BSM corrections to the $\gamma\ell\ell$ vertex. The Standard Model contribution to $d_{\ell}$ only appears at the 3-loop level, which for the $\tau$ lepton is $\left|d_{\tau}\right|\lesssim 10^{-34}\,e\cdot$cm~\cite{Hoogeveen:1990cb}.
The EDM of the $\tau$ lepton was measured by the Belle experiment using $e^{+}e^{-}\rightarrow\tau^{+}\tau^{-}$ events, yielding limits at the 95\% confidence level~\cite{Belle:2021ybo}: 
\begin{align}
-1.85\times10^{-17} & <\text{Re}\left\{ d_{\tau}\right\} <0.61\times10^{-17}\,e\cdot\text{cm}\,,\nonumber\\
-1.03\times10^{-17} & <\text{Im}\left\{ d_{\tau}\right\} <0.23\times10^{-17}\,e\cdot\text{cm}\,.
\end{align}

Lepton dipole moments are formally defined in the limit where the photon which the lepton couples to has zero momentum transfer (see Eq.~(\ref{Eqn:VertexParameterization}) and Eq.~(\ref{Eqn:dtauDefinition})). Since the Belle measurement was performed using an $s$-channel process with a momentum transfer of 10.58 GeV, a caveat here is that it is a measurement of $F_{3}\left(q^{2}=10.58\, \mathrm{GeV}\right)$ rather than the EDM.

In contrast to Belle, the quasi-real photons in the peripheral proton-proton collisions exploited by CMS enable a measurement of the $\tau$ EDM at almost zero momentum transfer. The current best direct upper limit on the $\tau$ EDM is~\cite{CMS:2024skm}:
\begin{equation}
    \left|d_{\tau}\right| < 2.9\times 10^{-17}\,e\cdot\text{cm}\,.\label{Eqn:DELPHIlimitsDtau}
\end{equation}

Given that $a_{\tau}$ and $d_{\tau}$ can be powerful probes of physics beyond the Standard Model, improving the measurement strategies using the capabilities of the LHC is important. With the heavy ion and proton programs of the LHC Run 3, precise measurements of the $\tau$ electromagnetic dipole moments by the LHC experiments are possible, surpassing the current best limits. The measurement improvements required for this can be realized not only by using larger datasets and reduced systematic uncertainties, but also by using experimental observables which are more sensitive than the ones used so far.

Optimal observables are kinematic functions that are most sensitive to a parameter of interest under certain assumptions. By design, they contain the maximum available information in the phase space, and can be used to extract the most stringent limits~\cite{Atwood:1991ka,Diehl:1993br}. They have been used in previous experimental analyses to probe triple gauge boson couplings~\cite{ALEPH:1997agc,L3:2004hlr,OPAL:1998ixj,OPAL:2000rnf,OPAL:2003xqq,OPAL:2003gfi}, CP~properties of Higgs boson couplings~\cite{ATLAS:2020evk,ATLAS:2022tan,ATLAS:2023mqy,CMS:2017len,CMS:2019jdw,CMS:2020cga,CMS:2021nnc}, searches for CP violation in $Z\rightarrow\tau\tau$ decays~\cite{OPAL:1994klt,OPAL:1996dwj}, in measurements of the polarization $\tau$ leptons produced in $Z\rightarrow\tau\tau$ decays~\cite{ALEPH:1993pdh,ALEPH:2001uca,L3:1994hzc,OPAL:1996krj}, and measurements of the electric dipole moment of the $\tau$ lepton~\cite{Belle:2002nla,Belle:2021ybo}. They have also been proposed as observables that can be used in flavour physics studies~\cite{Calcuttawala:2017usw}, measurements of anomalous top quark couplings~\cite{Hioki:2012vn}, etc. 

These advanced constructs can be used when the parameters of interest appear as multiplicative constants in the differential cross section of a process\footnote{Optimal observables cannot be used to measure masses and widths of particles which appear as additive constants, e.g. in propagators.}. Since lepton AMDM's and EDM's by definition are directly related to the photon-lepton coupling strength, optimal observables are an interesting technique to measure $a_{\ell}$ and $d_{\ell}$ in processes with $\gamma\ell\ell$ vertices. In this work, we study the potential use of optimal observables in improving existing measurements of $a_{\tau}$ in ultra-peripheral PbPb collisions at the LHC, and comment on the limitations for the use of optimal observables for $d_{\tau}$.

This paper is organized as follows: in Sec.~\ref{Sec:Theory} we outline our theoretical framework, including a description of optimal observables; in Sec.~\ref{Sec:MCSetup} our Monte Carlo simulations are described; in Sec.~\ref{Sec:OOstudyAtau} we study the sensitivity of optimal observables for the measurement of $a_{\tau}$; the challenges and potential implementations of optimal observables in experimental analyses are highlighted in Sec.~\ref{Sec:Discussion}; finally, we conclude in Sec.~\ref{Sec:Conclusion}.

\section{Theoretical and statistical framework}\label{Sec:Theory}
\subsection{\texorpdfstring{$\tau^{+}\tau^{-}$}{Di-tau} production in ultra-peripheral PbPb collisions}\label{SubSec:EMDipoleMoments}

Ultra-relativistic Pb ions, such as those colliding at the LHC, are accompanied by a large flux of quasi-real photons. The photons from two colliding Pb ions can interact and produce various final states with leptons, mesons, $W$ bosons, etc. Collisions with an impact parameter of $b\gtrsim 2 R_{\text{Pb}}$, with $R_{\text{Pb}}$ being the radius of the colliding lead ions, are denoted as ultra-peripheral collisions. Various experimental techniques can be used to isolate this type of collisions, such as vetoes on total transverse energy, activity in forward calorimeters, etc. In these clean environments, it is possible to study exclusive di-photon initiated processes.

In this work, we are interested in the following process:
\begin{equation}
\text{Pb}+\text{Pb}\rightarrow\text{Pb}+\text{Pb}+\gamma\gamma\rightarrow\text{Pb}+\text{Pb}+\tau^{+}\tau^{-}\,.\label{Eqn:Process}
\end{equation}

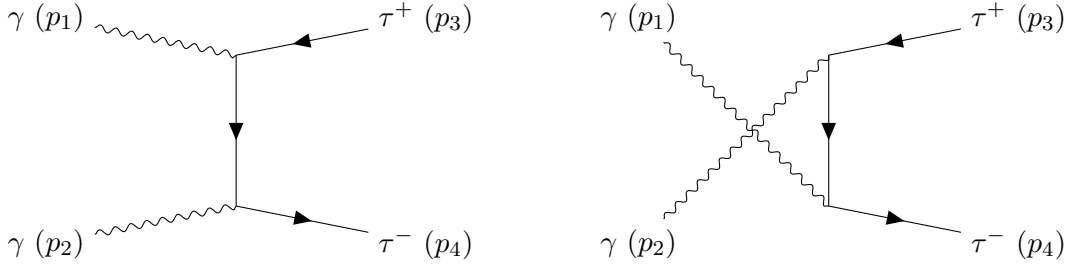
\begin{figure}[t]
    \centering
    \subfloat{
        \begin{tikzpicture}
            \begin{feynman} 
                \vertex (v1) at ( 0.0,+1.0); 
                \vertex (v2) at ( 0.0,-1.0);
                \vertex (y1) at (-2.5,+1.5) {$\gamma\,\left(p_{1}\right)$};
                \vertex (y2) at (-2.5,-1.5) {$\gamma\,\left(p_{2}\right)$};
                \vertex (t1) at (+2.5,+1.5) {$\tau^{+}\,\left(p_{3}\right)$};
                \vertex (t2) at (+2.5,-1.5) {$\tau^{-}\,\left(p_{4}\right)$};
                \diagram* {   
                    (t1) -- [fermion] (v1) -- [fermion] (v2)  -- [fermion] (t2);
                    (y1) -- [photon] (v1); 
                    (y2) -- [photon] (v2); 
                };
            \end{feynman}
    \end{tikzpicture}
    }
    \qquad\quad
    \subfloat{
        \begin{tikzpicture}
            \begin{feynman} 
                \vertex (v1) at ( 0.0,+1,0); 
                \vertex (v2) at ( 0.0,-1.0);
                \vertex (y1) at (-2.5,+1.5) {$\gamma\,\left(p_{1}\right)$};
                \vertex (y2) at (-2.5,-1.5) {$\gamma\,\left(p_{2}\right)$};
                \vertex (t1) at (+2.5,+1.5) {$\tau^{+}\,\left(p_{3}\right)$};
                \vertex (t2) at (+2.5,-1.5) {$\tau^{-}\,\left(p_{4}\right)$};
                \diagram* {   
                    (t1) -- [fermion] (v1) -- [fermion] (v2)  -- [fermion] (t2);
                    (y1) -- [photon] (v2); 
                    (y2) -- [photon] (v1); 
                };
            \end{feynman}
            \end{tikzpicture}
    }
    \caption{Feynman diagrams showing the leading-order QED contributions to the process shown in Eq.~(\ref{Eqn:Kinematics}).}\label{Fig:FeynDia}
\end{figure}

We define the kinematics of the elementary process as follows:
\begin{equation}
\gamma\left(p_{1};\lambda_{1}\right)+\gamma\left(p_{2};\lambda_{2}\right)\rightarrow\tau^{+}\left(p_{3};\lambda_{3}\right)+\tau^{-}\left(p_{4};\lambda_{4}\right)\,,\label{Eqn:Kinematics}
\end{equation}
where $p_{i}$ are the four-momenta of the particles and $\lambda_{i}$ are their helicities. We use the following definitions of $p_{i}$ in the $\gamma\gamma$ center-of-mass frame (assuming that the initial state photons are real and have zero transverse momentum): $p_{1}=\left(E,E\hat{z}\right)$, $p_{2}=\left(E,-E\hat{z}\right)$, $p_{3}=\left(E,E\beta\hat{p}\right)$ and $p_{4}=\left(E,-E\beta\hat{p}\right)$, where $\beta=\sqrt{1-m_{\tau}^{2}/E^{2}}$ is the velocity of the $\tau$~lepton, and $\hat{p}=\left(\sin\theta\cos\phi,\sin\theta\sin\phi,\cos\theta\right)$ gives the $\tau$ direction of flight.

The general structure of the $\gamma\ell\ell$ vertex contained within the process which satisfies gauge invariance is given by:
\begin{equation}
i\Gamma^{\mu}\left(p^{\prime},p\right)=-ie\left[\gamma^{\mu}F_{1}\left(q^{2}\right)+\frac{i}{2m_{\tau}}\sigma^{\mu\nu}q_{\nu}F_{2}\left(q^{2}\right)+\frac{1}{2m_{\tau}}\sigma^{\mu\nu}\gamma^{5}q_{\nu}F_{3}\left(q^{2}\right)\right]\,,\label{Eqn:VertexParameterization}
\end{equation}
where $p$ and $p^{\prime}$ are the four-momenta of the two fermions, $q=p^{\prime}-p$ is the four-momentum of the involved photon, $\sigma^{\mu\nu}=\frac{i}{2}\left[\gamma^{\mu},\gamma^{\nu}\right]$, and $F_{1,2,3}\left(q^2\right)$ are momentum-transfer dependent form factors. In the limit where the photon is real, i.e. $q^{2}\rightarrow 0$, these form factors define the electromagnetic dipole moments of the $\tau$ lepton. In particular:
\begin{align}
    F_{1}\left(0\right) &= 1 \,,\\
    F_{2}\left(0\right) &= a_{\ell}\,,\\
    F_{3}\left(0\right) &= d_{\ell}\frac{2m_{\ell}}{e} \,.\label{Eqn:dtauDefinition}
\end{align}
Using this parameterization, the matrix element for the interaction shown in Eq.~(\ref{Eqn:Kinematics}) is:
\begin{align}
i\mathcal{M}^{\lambda_{1},\lambda_{2},\lambda_{3},\lambda_{4}} & =\epsilon_{1\mu}\left(\lambda_{1}\right)\epsilon_{2\nu}\left(\lambda_{2}\right)\bar{u}\left(p_{3};\lambda_{3}\right)\left[i\Gamma^{\mu}\left(p_{3},p_{t}\right)\frac{i\left(\s{p_{t}}+m_{\ell}\right)}{p_{t}^{2}-m_{\ell}^{2}}i\Gamma^{\nu}\left(p_{t},-p_{4}\right)\right.\nonumber\\
 & \,\,\,\,\left.+i\Gamma^{\nu}\left(p_{3},p_{u}\right)\frac{i\left(\s{p_{u}}+m_{\ell}\right)}{p_{u}^{2}-m_{\ell}^{2}}i\Gamma^{\mu}\left(p_{u},-p_{4}\right)\right]v\left(p_{4},\lambda_{4}\right)\,,\label{Eqn:MatrixElement}
\end{align}
where $p_{t}=p_{3}-p_{1}=p_{2}-p_{4}$ and $p_{u}=p_{1}-p_{4}=p_{3}-p_{2}$ are the momenta of the internal $\tau$ leptons in the $t-$ and $u-$channel diagrams respectively.

Our parameters of interest are $a_{\tau}$ and $d_{\tau}$. Since $d_{\tau}$ is typically expressed in units of $e\cdot cm$, we instead use for convenience the following dimensionless parameter of interest:
\begin{equation}
    \tilde{d}_{\tau}=F_{3}\left(0\right)=d_{\tau}\frac{2m_{\tau}}{e}\,.
\end{equation} 
Using $a_{\tau}$ and $\tilde{d}_{\tau}$, Eq.~(\ref{Eqn:MatrixElement}) can be factorized as:
\begin{equation}
    \mathcal{M}^{\lambda_{1},\lambda_{2},\lambda_{3},\lambda_{4}}=\sum_{i,j=0}^{2}\left(a_{\tau}\right)^{i}\left(\tilde{d}_{\tau}\right)^{j}\mathcal{M}^{\lambda_{1},\lambda_{2},\lambda_{3},\lambda_{4}}_{ij}\,,\label{Eqn:FactorizedMatrixElement}
\end{equation}
where $\mathcal{M}_{00}$ defines the SM matrix element.

\subsection{Optimal Observables}\label{SubSec:OptimalObservables}

The standard approach in particle physics when attempting to measure a model parameter using experimental data is to employ an \textit{observable} sensitive to the parameter of interest (POI). Both, the shape and normalization of the distribution of the observable, might change with the POI. By comparing theoretical predictions for different values of the POI to measured data, confidence intervals on the POI can be extracted. 

Often, the goal of such an experimental analysis is to measure the POI as precisely as possible, in other words, to obtain the smallest confidence intervals. For certain kinds of POI's, such as coupling strengths in a Lagrangian, this can be achieved using \textit{optimal observables}. The framework of these observables is described below. The notation here loosely follows that of Ref. \cite{Diehl:1993br}.

We consider an arbitrary process whose matrix element is at most quadratic in the POI's $\alpha_{i}$. The cross section of the process, differential in some phase space variable $\phi$, can be expressed as the sum of cross section terms $S_i$, with $i=0,1,2$, indicating the associated order of the POI's:
\begin{equation}
    \frac{d\sigma}{d\phi} = S_{0}+\sum_{i}\alpha_{i}S_{1,i}+\sum_{i,j}\alpha_{i}\alpha_{j}S_{2,ij}\,.\label{Eqn:OO_DiffXsec}
\end{equation}

Given a fixed integrated luminosity $\mathcal{L}_\text{int}$ and the goal to measure $\alpha_{i}$ as accurately as possible, consider a set of unbiased estimators for $\alpha_i$, denoted as $\hat{\alpha}_{i}$, such as the estimators of the maximum likelihood method in the limit of large numbers. The expectation values for these estimators are $E\left[\hat{\alpha}_{i}\right]=\alpha_{i}$. 

Considering only first order contributions in $\alpha_i$, the set of estimators $\hat{\alpha}_{i}$ which minimizes the uncertainty on the measurements of $\alpha_{i}$ fulfills the relation\footnote{This holds when a sensitivity of the cross section to $\alpha_i$ is present in addition to the differential dependence.}~\cite{Diehl:1993br}:
\begin{equation}
    V_{i,j}^{-1} = \text{Cov}\left[\hat{\alpha}_{i},\hat{\alpha}_{j}\right]^{-1}=\mathcal{L}_{\text{int}} \int d\phi \frac{S_{1,i}S_{1,j}}{S_{0}}\,,\label{Eqn:OO_Variance}
\end{equation}
with $V_{i,j} = \text{Cov}\left[\hat{\alpha}_{i},\hat{\alpha}_{j}\right]$ indicating the (co-)variance matrix for the POI's $\alpha_i$.
Assume that the estimators $\hat{\alpha}_i$ are based on functions $\mathcal{O}_{i}(\phi)$ of the phase space variables $\phi$, i.e. $\hat{\alpha}_i=\hat{\alpha}_i(\mathcal{O}_{i}(\phi))$, such that $E\left[\sum_k\mathcal{O}_{i}\right]=E_0\left[\sum_k\mathcal{O}_{i}\right]+\sum_j c_{ij}\alpha_{j}$, with $E_0\left[\sum_k\mathcal{O}_{i}\right]$ being the expectation value without contributions from $\alpha_i$, coefficients $c_{ij}$ specifying the relation to the $\alpha_i$, and index $k$ running over the number of events. In this case, the covariance given by: 
\begin{equation}
    \text{Cov}\left[\hat{\alpha}_{i},\hat{\alpha}_{j}\right]=\frac{1}{c_{ij}}\cdot\text{Cov}\left[\sum_k\mathcal{O}_{i}(\phi_k),\sum_k\mathcal{O}_{j}(\phi_k)\right]\frac{1}{c_{ji}} \,,\label{Eqn:OO_Variance2}
\end{equation}
fulfills Eq.~(\ref{Eqn:OO_Variance}) automatically if $\mathcal{O}_{i}=S_{1,i}/S_{0}$. In this case, the uncertainty on the measurement of $\alpha_i$ becomes minimal. These functions $\mathcal{O}_{i}$ are referred to as \textit{optimal observables}. 

In analyses where the POI's $\alpha_i$ are measured independently of each other, i.e. when $S_{2,ij}=0$ for $i\neq j$ in Eq.~(\ref{Eqn:OO_DiffXsec}), one can generalize and define the first-order and second-order optimal observables as:
\begin{equation}
    \mathcal{OO}_{1,i}=\frac{S_{1,i}}{S_{0}}\qquad;\qquad\mathcal{OO}_{2,i}=\frac{S_{2,ii}}{S_{0}}\,,
\end{equation}
where the index $i$ means that the observables have been defined for the POI $\alpha_{i}$. The second-order optimal observable generally yields weaker constraints on $\alpha_{i}$ than the first-order optimal observable.

Since the matrix element squared of a scattering process and the differential cross section, are proportional to each other up to a phase space factor, the above arguments also hold when one uses $\left|\mathcal{M}^{2}\right|$ factorized in the same way as in Eq.~(\ref{Eqn:OO_DiffXsec}) to define the optimal observables.

In order to use the optimal observable technique to measure the $\tau$ AMDM, we can use Eq.~(\ref{Eqn:FactorizedMatrixElement}), with $\tilde{d}_{\tau}=0$. We have - with $\lambda_{1,2,3,4}$ indicating the involved helicities:
\begin{equation}
    \left|\mathcal{M}^{\lambda_{1},\lambda_{2},\lambda_{3},\lambda_{4}}\right|^{2}=\left|\mathcal{M}_{00}^{\lambda_{1},\lambda_{2},\lambda_{3},\lambda_{4}}\right|^{2}+a_{\tau}\cdot2\text{Re}\left\{ \left(\mathcal{M}_{00}^{\lambda_{1},\lambda_{2},\lambda_{3},\lambda_{4}}\right)^{*}\mathcal{M}_{10}^{\lambda_{1},\lambda_{2},\lambda_{3},\lambda_{4}}\right\} +\dots\,,\label{Eqn:MatrixElementSqFactorizedAtau} 
\end{equation}
where the higher order terms can be neglected for sufficiently small $a_{\tau}$. Since helicities are not directly accessible experimentally, averaging over particle helicities in Eq.~(\ref{Eqn:MatrixElementSqFactorizedAtau}) is required. Depending on which particle helicities are summed over, we define the following optimal observables:
\begin{align}
\ooatau & =\frac{\sum_{\lambda_{1},\lambda_{2},\lambda_{3},\lambda_{4}}2\text{Re}\left\{ \left(\mathcal{M}_{00}^{\lambda_{1},\lambda_{2},\lambda_{3},\lambda_{4}}\right)^{*}\mathcal{M}_{10}^{\lambda_{1},\lambda_{2},\lambda_{3},\lambda_{4}}\right\} }{\sum_{\lambda_{1},\lambda_{2},\lambda_{3},\lambda_{4}}\left|\mathcal{M}_{00}^{\lambda_{1},\lambda_{2},\lambda_{3},\lambda_{4}}\right|^{2}}\,,\label{Eqn:Formula_HAOO_atau}\\
\ooatauphavg & =\frac{\sum_{\lambda_{1},\lambda_{2}}2\text{Re}\left\{ \left(\mathcal{M}_{00}^{\lambda_{1},\lambda_{2},\lambda_{3},\lambda_{4}}\right)^{*}\mathcal{M}_{10}^{\lambda_{1},\lambda_{2},\lambda_{3},\lambda_{4}}\right\} }{\sum_{\lambda_{1},\lambda_{2}}\left|\mathcal{M}_{00}^{\lambda_{1},\lambda_{2},\lambda_{3},\lambda_{4}}\right|^{2}}\,,\label{Eqn:Formula_PHAOO_atau}
\end{align}
where the $\lambda_i$ superscripts indicate the helicities that are assumed to be known for the given event. The \ooatauphavg distribution contains four distinct sub-distributions depending on the choice of $\lambda_{3,4}$. In order to compare the difference in sensitivity to $a_{\tau}$ with the hypothetical case where all particle helicities are known, we also define the following optimal observable:
\begin{equation}
    \ooataufullpol=\frac{2\text{Re}\left\{ \left(\mathcal{M}_{00}^{\lambda_{1},\lambda_{2},\lambda_{3},\lambda_{4}}\right)^{*}\mathcal{M}_{10}^{\lambda_{1},\lambda_{2},\lambda_{3},\lambda_{4}}\right\} }{\left|\mathcal{M}_{00}^{\lambda_{1},\lambda_{2},\lambda_{3},\lambda_{4}}\right|^{2}}\,.\label{Eqn:Formula_FPOO_atau}
\end{equation}

Eq.~(\ref{Eqn:Formula_FPOO_atau}) can give 16 possible values of the observable for a given phase space point, depending on the choice of $\lambda_{1,2,3,4}$. We refer to Eq.~(\ref{Eqn:Formula_FPOO_atau}) as the \textit{fully polarized} optimal observable for $a_{\tau}$, while Eq.~(\ref{Eqn:Formula_PHAOO_atau}) and Eq.~(\ref{Eqn:Formula_HAOO_atau}) are denoted as the \textit{photon helicity averaged} and \textit{helicity averaged} optimal observables for $a_{\tau}$, respectively. 

It is important to note the symmetry $\mathcal{M}^{\lambda_{1},\lambda_{2},\lambda_{3},\lambda_{4}}=\mathcal{M}^{-\lambda_{1},-\lambda_{2},-\lambda_{3},-\lambda_{4}}$, which is a direct consequence of parity conservation in the electromagnetic interaction. Therefore, we find that:
\begin{equation}
    \sum_{\lambda_{1},\lambda_{2},\lambda_{3}}\left|\mathcal{M}^{\lambda_{1},\lambda_{2},\lambda_{3},+}\right|^{2} = \sum_{\lambda_{1},\lambda_{2},\lambda_{3}}\left|\mathcal{M}^{\lambda_{1},\lambda_{2},\lambda_{3},-}\right|^{2}\,.
\end{equation}
Both the quantities in the above equation appear in $\sum_{\lambda_{1},\lambda_{2},\lambda_{3},\lambda_{4}}\left|\mathcal{M}^{\lambda_{1},\lambda_{2},\lambda_{3},\lambda_{4}}\right|^{2}$. Thus, the optimal observable computed in the case, where only one $\tau$ helicity is known, is equal (up to a proportionality constant) to the optimal observable computed when all four helicities are summed over. A statistical analysis will yield the same confidence intervals on $a_{\tau}$, so it is not necessary to consider the observable where only one $\tau$ helicity is known. Note that this relation holds as well for the hypothetical case when three of the four helicities are known, and only one helicity is summed over.

Using the center-of-mass (COM)  kinematics defined in Sec.~\ref{SubSec:EMDipoleMoments}, the helicity averaged optimal observable for $a_{\tau}$ has the simple expression:
\begin{align}
\ooatau & =\frac{4\left(\beta^{2}\cos^{2}\theta-1\right)}{\beta^{4}\left(1+\sin^{4}\theta\right)-2\beta^{2}\sin^{2}\theta-1}\,.\label{Eqn:AnalyticExpressionAtauOO}
\end{align}

We can, in principle, define similar observables for the electric dipole moment of the $\tau$ lepton. Considering only the longitudinal spin components of the particles involved, i.e. their helicities, denoted as the \emph{helicity amplitude framework}, the term linear in $\tilde{d}_{\tau}$, $\mathcal{M}_{01}^{\lambda_{1},\lambda_{2},\lambda_{3},\lambda_{4}}$, however, vanishes for all combinations of particle helicities except for the $\pm\pm\mp\mp$ ones. These helicity combinations account for $\mathcal{O}\left(10^{-3}\,\%\right)$ of all events in our signal regions (see Sec.~\ref{subsec:fiducialselection}), with very small variation depending on the value of $\tilde{d}_{\tau}$. Since this amounts to an exceedingly small number of events an experiment would potentially see, it is essentially impossible to use first-order optimal observables in the helicity amplitude framework for measuring $\tilde{d}_{\tau}$. Second-order optimal observables are possible to define in the helicity amplitude framework, but by design are only sensitive to the absolute value of $\tilde{d}_{\tau}$. Any constraints obtained from these will be weaker than for first-order optimal observables. 

Recently, the matrix element of the $\gamma\gamma\rightarrow\tau^{+}\tau^{-}$ process has been computed with full $\tau$-spin correlations, averaged over the polarizations of the photons~\cite{Banerjee:2023qjc}, where it was found that a term in the $\left|\mathcal{M}\right|^{2}$ linear in $\tilde{d}_{\tau}$ exists due to transverse spin correlations between the $\tau^{+}$ and $\tau^{-}$. When the transverse components of the $\tau$ spin vectors are integrated over, the term linear in $\tilde{d}_{\tau}$ term vanishes. This action reduces the 3-dimensional spin degree-of-freedom to the one-dimensional helicity degree-of-freedom, thereby confirming our observation that only a second order optimal observable for $\tilde{d}_{\tau}$ can be defined in the helicity amplitude framework. The calculations presented in Ref.~\cite{Banerjee:2023qjc} have been implemented as a reweighting tool for $\tau$ decays in an update of the \texttt{TauSpinner} package~\cite{Czyczula:2012ny}, and can in principle be used to study the sensitivity of a first-order optimal observable for $\tilde{d}_{\tau}$. Such a phenomenological analysis, however, requires that the full $\tau$ spin vectors are written to the event files, which is not supported in current Monte Carlo infrastructure, and is therefore beyond the scope of this paper. 

\subsection{Statistical analysis}\label{Subsec:StatisticalAnalyis}

To extract confidence intervals on $a_{\tau}$ we use an extended binned maximum likelihood fit~\cite{cowan1998statistical}. 
Considering a histogram of an observable sensitive to only one parameter of interest $\alpha$, with $n_{i}$ events in the $i^{\text{th}}$ bin, the likelihood is constructed assuming that the events are Poisson distributed:
\begin{equation}
    L\left(\alpha|\{n_{i}\}\right)=\prod_{i}\frac{\left(\mu_{i}\left(\alpha\right)\right)^{n_{i}}e^{-\mu_{i}\left(\alpha\right)}}{n_{i}!}\,,
\end{equation}
where $\mu_{i}\left(\alpha\right)$ is the expected number of events in the $i^{\text{th}}$ bin for the given $\alpha$ value. In our work, we use as data $n_{i}$ the predictions for the null hypotheses $a_{\tau}=0$ which corresponds to the tree level QED predictions, i.e. $n_{i}=\mu_{i}(a_{\tau}=\tilde{d}_{\tau}=0)$.

We compute the negative logarithm of the likelihood function and shift it to the minimum, denoted as the delta negative log-likelihood ($\Delta$NLL), to which we fit a polynomial of degree $N$ close to $a_{\tau}=0$. We consider histograms of the transverse momentum ($p_{T}$) of the muon from one of the $\tau$-lepton decays, as well as the three defined optimal observables. For the muon $p_{T}$ and both helicity averaged optimal observables, $N=8$ produces a description of the $\Delta$NLL curve that is numerically stable and has sufficiently small residuals. For the optimal observables where some (or all) particle helicities are known, $N=10$ was found to be suitable. The 68\% and 95\% confidence intervals are determined from the $\Delta$NLL curve, and correspond to the $a_{\tau}$ values where the curve crosses the values 0.5 and 1.96 respectively.

Our analysis is performed using only the signal $\gamma\gamma\rightarrow\tau^{+}\tau^{-}$ events at truth-level, i.e. without a detector simulation, background predictions, and considering statistical uncertainties only. Since we are primarily interested in comparing the sensitivities to $a_{\tau}$ of different observables, we consider this approach sufficient. Comparisons to existing measurements should, however, be taken with a large grain of salt. Backgrounds in the ATLAS measurement based on ultra-peripheral PbPb collisions~\cite{ATLAS:2022ryk} amount to about 7\%-16\% of the observed events, depending on the event selection, and both, ATLAS and CMS measurements using ultra-peripheral PbPb collisions~\cite{ATLAS:2022ryk,CMS:2022arf} are still limited by low statistics. Ideally, the extraction of parameters of interest at truth-level, i.e. without detector simulation, should provide the same results as an extraction at detector-level, i.e. including a detector simulation, but impacts cannot be excluded. In comparison to existing measurements our results must, thus, be considered as strongly optimistic, though judgement on the relative performance of observables -- which we are interested in -- should be unaffected.

\section{Event simulation}\label{Sec:MCSetup}

We use the following simulated event samples in the analysis. We use \texttt{gamma-UPC} \cite{Shao:2022cly} to simulate the photon flux associated with colliding Pb ions. The Charged Form Factor option is chosen for the photon flux which -- as described in Ref.~\cite{Shao:2022cly} -- uses the realistic Woods-Saxon potential to simulate the charge distribution of the Pb ions. \texttt{MadGraph5\_aMC@NLO}~ \cite{Alwall:2014hca} is used to simulate the matrix elements and phase space integration of the $\gamma\gamma\rightarrow\tau^{+}\tau^{-}$ process, and \texttt{Pythia v8.245} \cite{Sjostrand:2014zea} to simulate the decays of the $\tau$ leptons. Final state QED radiation is simulated using \texttt{Photos v3.61} \cite{Davidson:2010ew}. The presence of a non-zero $a_{\tau}$ and $\tilde{d}_{\tau}$ as defined in Eq.~(\ref{Eqn:MatrixElement}) are implemented using a custom \texttt{Universal FeynRules Output} \cite{Degrande:2011ua} file generated using \texttt{FeynRules v2.3} \cite{Alloul:2013bka}.

\subsection{Matrix element reweighting}\label{SubSec:Reweighting}

Rather than producing dedicated simulated event samples for each assumed $a_{\tau}$ value, we employ matrix element reweighting to save computational resources. In this technique, a large sample is produced for a reference $a_{\tau}$ value. Then, the Monte Carlo weight~$w$ associated with each event~$i$ is modified using the following equation:
\begin{equation}
    w_{\text{new}}^{i} = \frac{\left|\mathcal{M}_{\text{new}}^{i}\right|^{2}}{\left|\mathcal{M}_{\text{old}}^{i}\right|^{2}} \cdot w_{\text{old}}^{i} = R^{i} \cdot w_{\text{old}}^{i}\,,
\end{equation}
where \textit{old} refers to the initial $a_{\tau}$ value and \textit{new} refers to a target $a_{\tau}$ value which is different from the initial one. In order to specify an uncertainty on an observable $\mathcal{O}$, such as the total cross-section, after the reweighting procedure, we define, following~\cite{Mattelaer:2016gcx}:
\begin{equation}
    \Delta\mathcal{O}_{\text{new}} = \left<R\right>\cdot\Delta\mathcal{O}_{\text{old}}+\Delta R \cdot \mathcal{O}_{\text{old}}\,,
    \label{eq:reweight-unc}
\end{equation}
where $\left<R\right>$ and $\Delta R$ are the mean and standard deviation, respectively, of the $R^i$ from the simulated events. We use helicity amplitude code automatically produced by \texttt{Madgraph5} via the \texttt{HELAS} library~\cite{Murayama:1992gi}, to compute the reweighting factors $R^i$.

\begin{figure}[t]
\centering
\subfloat[]{\includegraphics[width=0.48\textwidth]{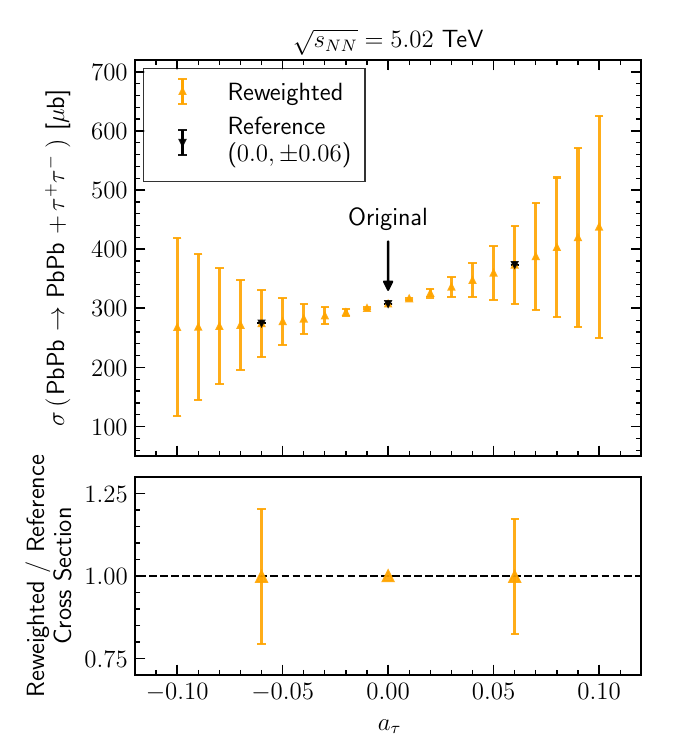}\label{subfig:rwgt_atau_0p0}}\hfill
\subfloat[]{\includegraphics[width=0.48\textwidth]{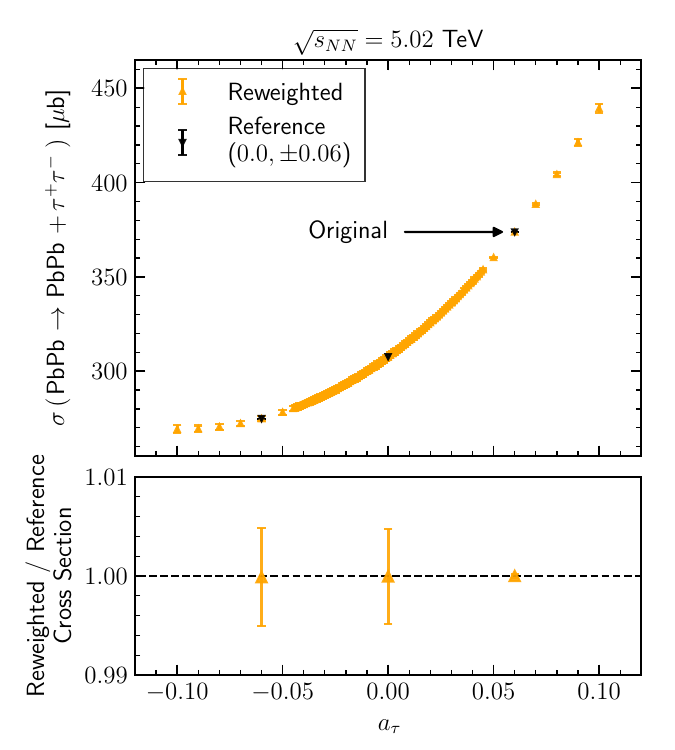}\label{subfig:rwgt_atau_p006}}\\
\caption{Total cross section of the $\text{PbPb}\rightarrow\text{PbPb}+\tau^{+}\tau^{-}$ process as a function of $a_{\tau}$, when performing the reweighting from an initial value of (a) $a_{\tau}=+0.0$ and (b) $a_{\tau}=+0.06$ (in the figures, \textit{Original} refers to the initial choice of $a_{\tau}$). The lower panels show the ratios of reweighted to reference cross sections.}\label{Fig:reweighting_xsec}
\end{figure}

In order to validate the matrix element reweighting procedure, we simulate $5$ million parton-level events for $a_{\tau}=0.0,\pm0.06$, and study the reweighting closure and the variation in the cross section uncertainty for different initial choices of $a_{\tau}$. The results of this validation are shown in Fig.~\ref{Fig:reweighting_xsec}. When starting with an initial choice of $a_{\tau}=0.0$ (see Fig.~\ref{subfig:rwgt_atau_0p0}), we observe that while the central value of the reweighted cross section is close to the reference cross section, the uncertainty blows up with increasing $|a_{\tau}|$. This is understood by the fact that since certain helicity combinations result in a zero matrix element for $a_{\tau}=\tilde{d}_{\tau}=0$ (see Eq.~(\ref{Eqn:MatrixElement})), the space of helicity combinations covered by an $a_{\tau}=0.0$ sample is a subset of that of an $a_{\tau}\neq0.0$ sample. Thus, reweighting from $a_{\tau}=0.0$ to $a_{\tau}\neq0.0$ incurs large uncertainties, which are dominated by $\Delta R$, i.e. the standard deviation of $R^{i}$. In contrast, reweighting in the other direction, i.e. from $a_{\tau}\neq0.0$ to $a_{\tau}=0.0$ does not incur such a cost (see Fig.~\ref{subfig:rwgt_atau_p006}). Starting with an initial choice of $a_{\tau}=+0.06$ yields $\mathcal{O}\left(0.5\%\right)$ uncertainties on the reweighted cross section in the full $a_{\tau}$ range of interest. Though the uncertainties in the total cross sections, calculated following Eq.~(\ref{eq:reweight-unc}), are not used further in the study, the reweighting with an initial choice of $a_{\tau}=+0.06$ is chosen as preferred.

Our analysis of the sensitivity of optimal observables to $a_{\tau}$ is thus performed based on the Monte Carlo sample generated for $a_{\tau}=+0.06$, reweighted from $a_{\tau}=-0.045$ to $a_{\tau}=+0.045$ in steps of $0.001$. This range is chosen to encompass the current limits on $a_{\tau}$ (Eq.~(\ref{Eqn:DELPHIlimitsAtau})). We reweight to $a_{\tau}=\pm0.05,\pm0.06,\pm0.07,\pm0.08,\pm0.09,\pm0.10$, in order to study the behaviour of the $\Delta$NLL curves for very large $a_{\tau}$ values. 

\subsection{Fiducial selection at particle level}\label{subsec:fiducialselection}

Our fiducial selection of particle level events closely follows the definitions of the ATLAS signal regions~\cite{ATLAS:2022ryk}. Final state particles are required to satisfy a pseudo-rapidity\footnote{The pseudo-rapidity $\eta$ is defined through the polar angle $\theta$ of a cylindrical coordinate system, with its origin at nominal collision point of the particles and the longitudinal axis along the colliding Pb ion direction, as $\eta=-\ln\tan(\theta/2)$.} cut of $\left|\eta\right|<2.5$. Leptons i.e. electrons and muons are required to have transverse momentum $p_{T}>4.0$ GeV, and charged hadrons, hereafter referred to as tracks, are required to have $p_{T}>0.1$ GeV. Using these objects, we define our signal regions:
\begin{itemize}
    \item 1 muon + 1 track ($1\mu1T$): We require exactly 0 electrons, 1 muon, and 1 track in the final state. The muon and track must be separated by $\sqrt{\Delta\eta^{2}+\Delta\phi^{2}}>0.1$, where $\phi$ is the azimuthal angle. The combined muon + track system must have a charge of zero, $p_{T}>1.0$ GeV and acoplanarity $A_{\phi}=1-\Delta\phi\left(\mu,\mathrm{trks}\right)/\pi<0.4$.
    \item 1 muon + 3 tracks ($1\mu3T$): We require exactly 0 electrons, 1 muon, and 3 tracks in the final state. The muon must be separated from each track by $\sqrt{\Delta\eta^{2}+\Delta\phi^{2}}>0.1$. The acoplanarity of the muon and 3-track system must be less than 0.2, the combined muon + 3 track system must have zero charge, and the invariant mass of the 3-track system must be less than $1.7$ GeV.
    \item 1 muon + 1 electron ($1\mu1e$): We require exactly 1 electron, 1 muon, and 0 tracks in the event. The electron and muon must have opposite charges, and are separated by $\sqrt{\Delta\eta^{2}+\Delta\phi^{2}}>0.1$.
\end{itemize}

At an integrated luminosity of 2.0 nb$^{-1}$, corresponding roughly to the integrated luminosity of PbPb collisions in LHC Run 2, the yields are 1591, 540, and 179 events for the $1\mu1T$, $1\mu3T$, and $1\mu1e$ signal regions, respectively for $a_{\tau},\tilde{d}_{\tau}=0,0$.

\section{Study of optimal observables for \texorpdfstring{$a_{\tau}$}{tau AMDM}}\label{Sec:OOstudyAtau}

\begin{figure}[t]
\centering
\subfloat[]{\includegraphics[width=0.43\textwidth]{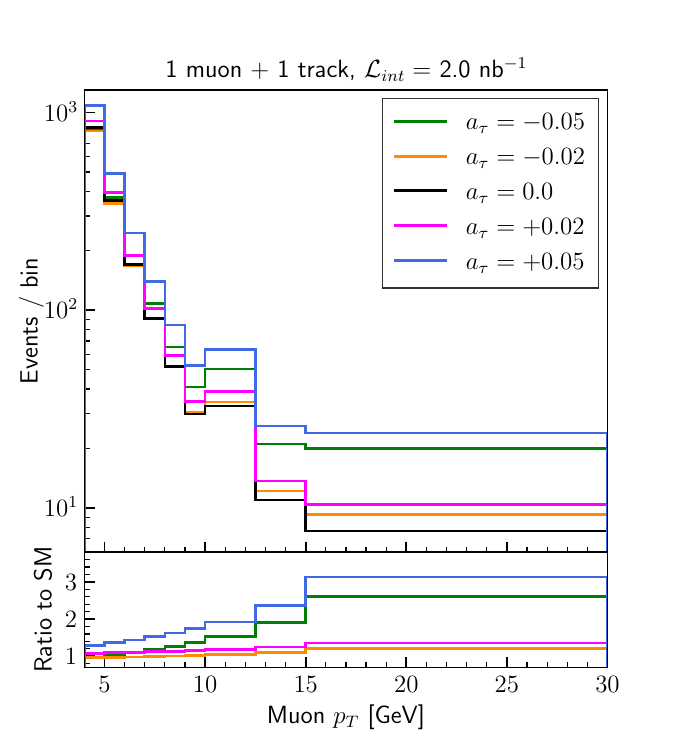}\label{subfig:DistribMuPt}}\hspace{1cm}
\subfloat[]{\includegraphics[width=0.43\textwidth]{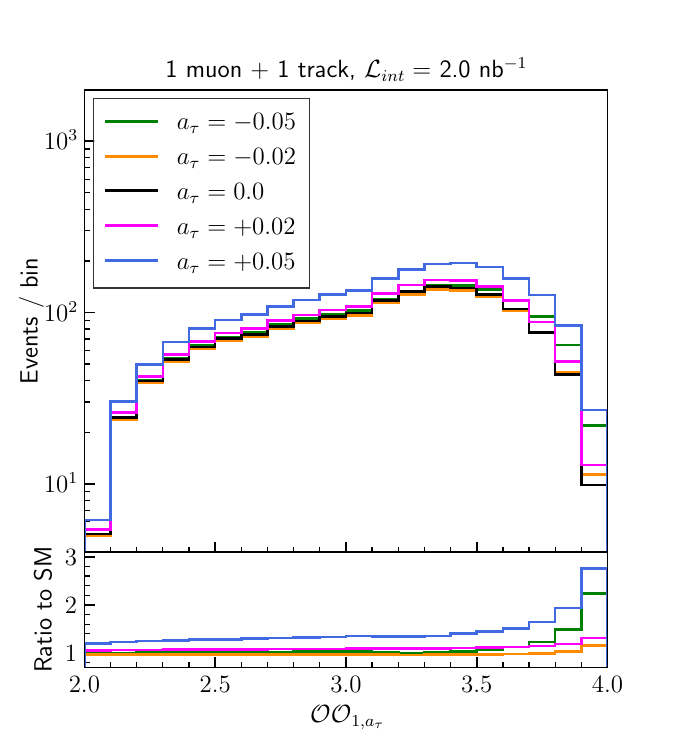}\label{subfig:DistribHAOO}}\\[-0.45cm]
\subfloat[]{\includegraphics[width=0.43\textwidth]{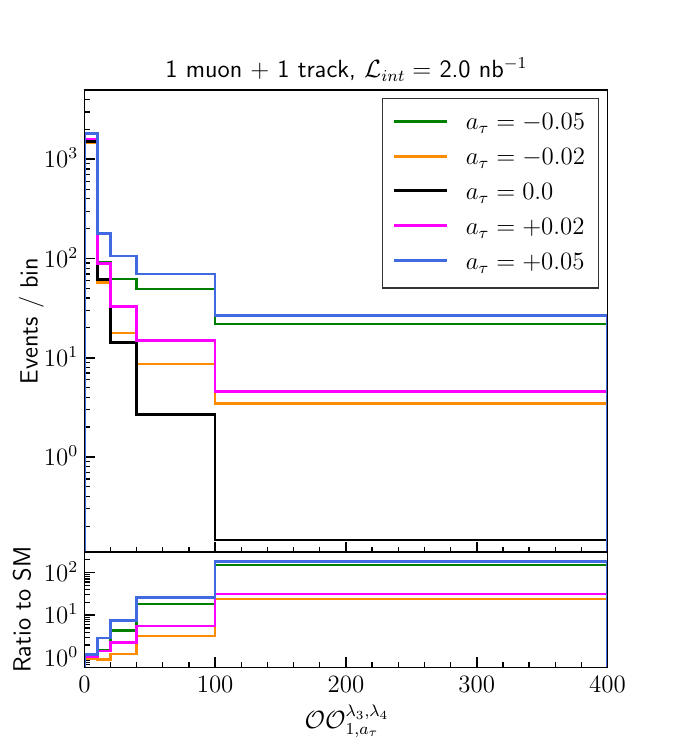}}\label{subfig:DistribPHAOO}\hspace{1cm}
\subfloat[]{\includegraphics[width=0.43\textwidth]{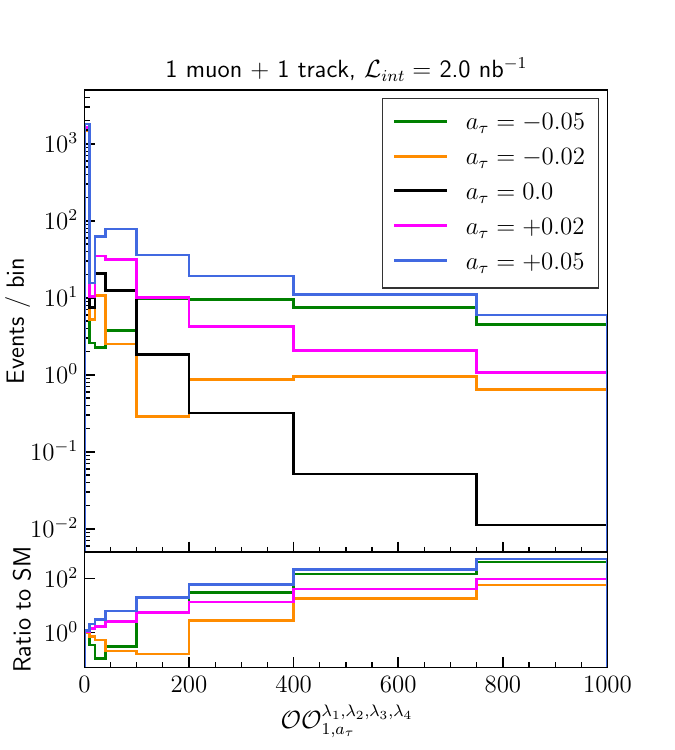}\label{subfig:DistribFPOO}}\\
\caption{Distributions of observables sensitive to $a_{\tau}$: (a) muon $p_{T}$, (b) \ooatau, (c) \ooatauphavg, and (d) \ooataufullpol, as defined in Eqs.~(\ref{Eqn:Formula_PHAOO_atau}), ~(\ref{Eqn:Formula_HAOO_atau}) and~(\ref{Eqn:Formula_FPOO_atau}), respectively. The observables are shown for different values of $a_{\tau}$, 
 with the lower panels indicating the ratios of the distributions to the SM prediction of $a_{\tau}=0.0$. Note the different scales of the ratios in the lower panels for the four observables.}\label{Fig:Distributions}
\end{figure}

We compare the limits on $a_{\tau}$ obtained from the optimal observables, with those obtained using the observable used in Ref.~\cite{ATLAS:2022ryk}, namely the $p_{T}$ of the muon from the muonic $\tau$ decay. We compute the optimal observables using the 4-vectors of the $\tau$ leptons from the hard interaction, boosted into the COM frame. We rely, thus, on Monte-Carlo simulation to determine the optimal observables -- which is possible in our truth-level study, but corresponds to an optimistic scenario in real life. With data, reconstructed 4-vectors of the $\tau$ leptons would have to be used which can only be determined approximately due to the loss of information from $\tau$-decay neutrinos escaping detection. 

We show the distributions of the optimal observables for the $1\mu 1T$ signal region, normalized to an integrated luminosity of $2.0$ nb$^{-1}$ in Fig.~\ref{Fig:Distributions}. The binning is chosen such as to provide sufficient statistics in each bin. The muon $p_{T}$ distribution is shown for comparison, where the same binning is used as in Ref.~\cite{ATLAS:2022ryk}. The lower panels in Fig.~\ref{Fig:Distributions} display the ratios of the predictions for non-zero $a_{\tau}$ values to that of the SM prediction. The size of ratios different from unity, reaching values of > 10 and even > 100 in cases where the particle helicities are known, demonstrates that optimal observables observables are highly sensitive to $a_{\tau}$.

\begin{table}[!t]
\begin{centering}
\begin{tabular*}{0.8\textwidth}{c @{\extracolsep{\fill}} c @{\extracolsep{\fill}} c}
\hline 
Observable & 68\% CI & 95\% CI\tabularnewline
\hline 
\hline 
Muon $p_{T}$ & $\left[-0.0088 , 0.0057\right]$ & $\left[-0.0225 , 0.0102\right]$\tabularnewline
\ooatau & $\left[-0.0090 , 0.0057\right]$ & $\left[-0.0280 , 0.0101\right]$\tabularnewline
\ooatauphavg & $\left[-0.0053 , 0.0031\right]$ & $\left[-0.0102 , 0.0057\right]$\tabularnewline
\ooataufullpol & $\left[-0.0020 , 0.0022\right]$ & $\left[-0.0043 , 0.0044\right]$ \tabularnewline
\hline 
\end{tabular*}
\par\end{centering}
\caption{Confidence intervals (CI's) obtained for $a_{\tau}$ using the muon $p_{T}$ and the optimal observables, as shown in Fig.~\ref{Fig:Distributions}. All three signal regions defined in Sec.~\ref{subsec:fiducialselection} were considered. Note we use the particle-level muon $p_{T}$, and that the optimal observables are computed using parton-level four-momenta and helicities. Moreover, the CI's are obtained using only the signal $\gamma\gamma\rightarrow\tau^{+}\tau^{-}$ process, without considering systematic uncertainties.}\label{table:LimitsAtau}
\end{table}

These distributions are used to extract confidence intervals (CI's) on $a_{\tau}$, using the procedure described in Sec.~\ref{Subsec:StatisticalAnalyis}. The obtained results are summarized in Tab.~\ref{table:LimitsAtau}. When fewer particle helicities are summed over, or, in other words, when more information about the event is known, the constraints on $a_{\tau}$ become stricter. Contradictory to expectation, however, the helicity-averaged optimal observable \ooatau yields slightly looser constraints than the muon $p_{T}$ observable, particularly for the 68\%-lower and 95\%-lower CI boundaries. The muon $p_{T}$ observable is implicitly also helicity-averaged, since the helicities of the $\tau$ leptons and photons are not considered when filling the distribution, it can be placed on the same footing as \ooatau. Seemingly, the promise of the optimal observable being truly optimal is under question.

As pointed out in Ref.~\cite{Diehl:1993br}, phase space cuts can impact the performance of optimal observables, to the degree that the expected behaviour, i.e. the optimal sensitivity to the parameter of interest is not observed. This effect is captured in Eq.~(\ref{Eqn:OO_Variance}) through the integration boundaries. A phase space cut restricts the domain of integration over the phase space variable $\phi$, thereby making the integral on the right hand side smaller. As a consequence, the inverse of the estimator variance on the left hand side, i.e. $V_{i,j}^{-1}$, reduces as well, or the variance $V_{i,j}$ increases. Thus, it is possible for simple kinematic observables, like muon $p_{T}$ in our case, to outperform the optimal observable. The expected behaviour can be restored either by enlarging the phase space, or by increasing the integrated luminosity $\mathcal{L}_{\text{int}}$, which corresponds to running the experiment for a longer time.

We tested the effect in the following manner. The $p_{T}$ requirement used to define the muon object at particle-level is decreased and increased to $3$ GeV and $5$ GeV, respectively. This amounts to looser and stricter phase space cuts, compared to the nominal $4$ GeV cut. The $3$ GeV cut increases the event yield by roughly a factor of 2, while the $5$ GeV cut decreases the event yield by a factor of 2. In addition, we scan over the assumed integrated luminosity, from 0.5 nb$^{-1}$ to 14.5 nb$^{-1}$. For each integrated luminosity, we compute the difference between the 68\% and 95\% CI boundaries of the muon $p_{T}$ and \ooatau observables. For the lower boundaries, we invert the sign of the difference, so that the interpretation remains the same for upper and lower interval boundaries: a positive difference implies that \ooatau outperforms the muon $p_{T}$. 

\begin{figure}[t]
\centering
\subfloat[]{\includegraphics[width=0.47\textwidth]{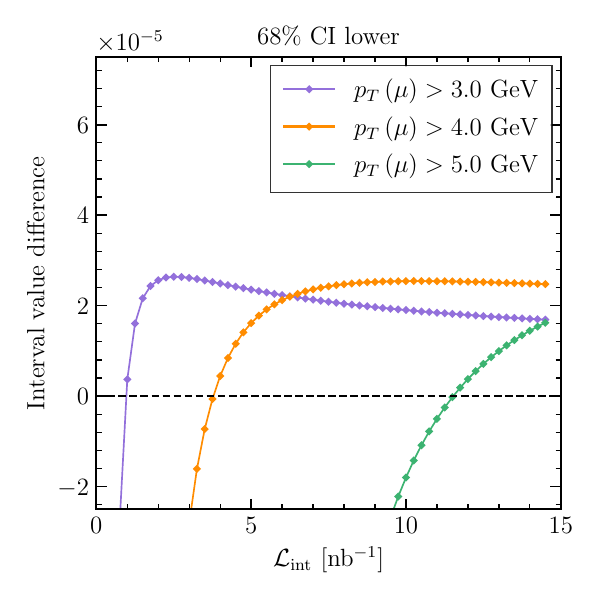}\label{subfig:68lowerstandard}}\hfill
\subfloat[]{\includegraphics[width=0.47\textwidth]{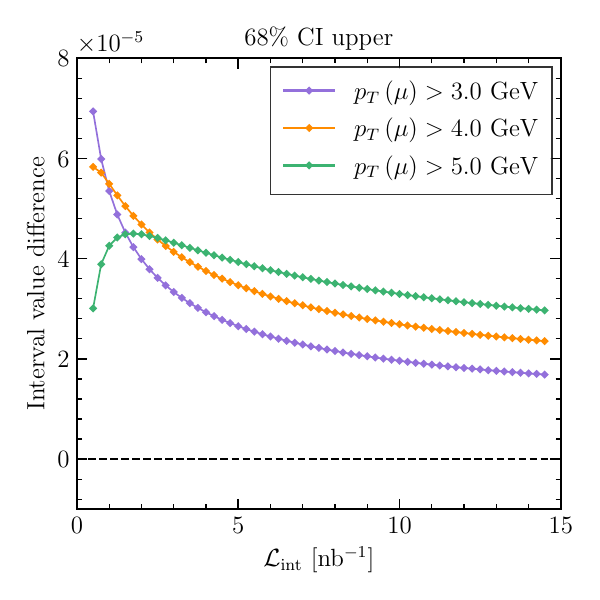}\label{subfig:68upperstandard}}\\
\caption{Difference between the (left) lower and (right) upper values of the 68\% confidence intervals (CI's), obtained for the muon $p_{T}$ and \ooatau observables. The different colors indicate the selection applied on the muon $p_{T}$ when defining the signal regions.}\label{Fig:68intervaldiff}
\end{figure}

The results of this study are given in Fig.~\ref{Fig:68intervaldiff}, showing the effect of phase space cuts and the integrated luminosity on the relative performance of the muon $p_{T}$ and \ooatau observables. While the optimal observable remains optimal when looking at the 68\%-upper value, the 68\%-lower value highlights the worse performance of \ooatau for part of the parameter space, including the default choices of $p_{T}\left(\mu\right)>4$~GeV and $\mathcal{L}_{\mathrm{int}}=2.0$~nb$^{-1}$ used in our analysis. If $\mathcal{L}_{\mathrm{int}}$ is increased, the optimal observable gains in sensitivity, eventually outperforming the muon $p_{T}$. The threshold where \ooatau improves over muon $p_{T}$ is at a lower $L_{\mathrm{int}}$ value the lower the muon $p_{T}$ requirement, i.e. the larger the phase space.

The same conclusions were obtained when modifying the requirements on other phase space variables, such as the $p_{T}$ cut used to define the track object. The observation still holds if the distributions of other kinematic observables, such as the $p_{T}$ of the track(s) and electrons in our signal regions, are used to obtain CI's on $a_{\tau}$, indicating that there is no bias due to the fact that we use the muon $p_{T}$ simultaneously as an observable and a variable to define our signal regions. The same observations were also made for the 95\% CI's instead of the 68\% CI's, where the effect of phase space cuts is more pronounced. 

\section{Further remarks on the optimal observables}\label{Sec:Discussion}
\subsection{Matrix element features}\label{SubSec:MatElemFeatures}

\begin{figure}[t]
\centering
\includegraphics[height=0.5\textwidth]{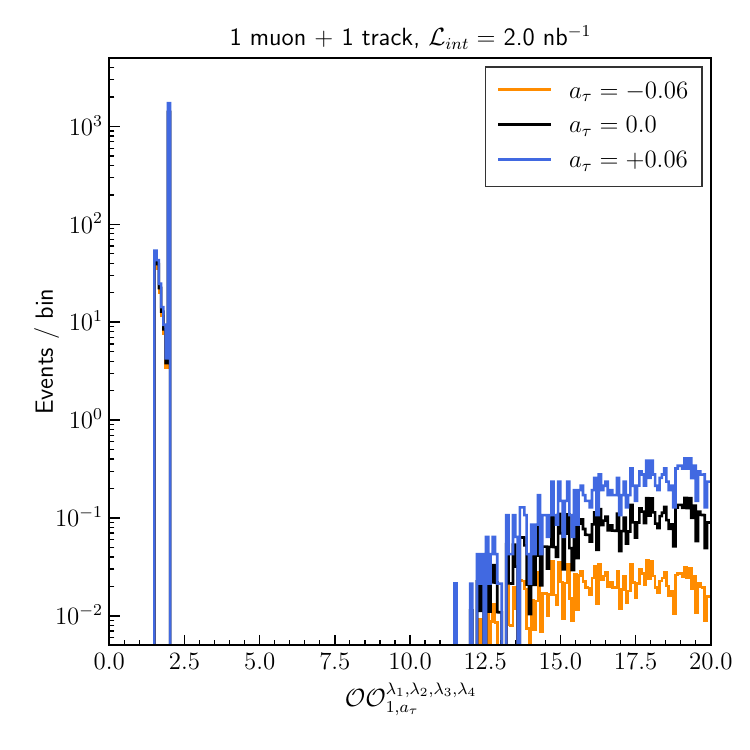}
\caption{Distribution of the \ooataufullpol observable, shown for different $a_{\tau}$ values, with finer binning compared to Fig.~\ref{subfig:DistribFPOO} and close to zero.}\label{Fig:DistribFPOOFineBin}
\end{figure}

The choice of binning for the optimal observables for $a_{\tau}$, as shown in Fig.~\ref{Fig:Distributions}, hides certain features of their distributions. Investigating in more detail, the following matrix elements needed for the fully polarized optimal observable \ooataufullpol can be computed, using the COM kinematics described in Sec.~\ref{SubSec:EMDipoleMoments}:
\begin{align}
    \mathcal{M}_{00}^{+-+-}&=\mathcal{M}_{10}^{+-+-}=\frac{4\beta\sin\theta\cos^{2}\frac{\theta}{2}}{\beta^{2}\cos^{2}\theta-1}\,,\nonumber\\
    \mathcal{M}_{00}^{+--+}&=\mathcal{M}_{10}^{+--+}=\frac{4\beta\sin\theta\sin^{2}\frac{\theta}{2}}{\beta^{2}\cos^{2}\theta-1}\,.
\end{align}
These matrix elements together with Eq.~(\ref{Eqn:Formula_FPOO_atau}) determine the fully polarized optimal observable to be $\ooataufullpol=2$ for the $\pm\mp\pm\mp$ and $\pm\mp\mp\pm$ helicity combinations, independent of the kinematics of the events. This results in a clear peak visible at the value of two in Fig.~\ref{Fig:DistribFPOOFineBin}, which displays the \ooataufullpol distribution with a very fine binning for values close to zero.
Further helicity combinations result in value ranges for \ooataufullpol, disjoint from those of other helicity combinations, leading to gaps in the distribution. As a consequence of these features, we had to use large binning in our analysis (see Fig.~\ref{subfig:DistribFPOO}) to ensure that no empty bins appear in the statistical analysis. These features were also seen in the distribution of \ooatauphavg, since the other contributing matrix elements give $\mathcal{M}^{\pm\pm+-}=\mathcal{M}^{\pm\pm-+}=0$. 

In an experimental analysis, the optimal observables for $a_{\tau}$ would have to be calculated from the kinematics of the $\tau$ decay products, by first estimating the phase space variables $\beta$ and $\theta$ of the $\tau$ leptons in the COM frame (see Eq.~(\ref{Eqn:AnalyticExpressionAtauOO}) for the helicity averaged optimal observable as function of $\beta$ and $\theta$).
One interesting possibility is to directly estimate the optimal observables (including the non-helicity-averaged ones) using advanced machine learning (ML) techniques like neural networks or symbolic regression~\cite{Butter:2021rvz} from the $\tau$ lepton decay products. In preliminary studies of these approaches for estimating \ooataufullpol we have observed that the $\pm\mp\pm\mp$ and $\pm\mp\mp\pm$ helicities -- having $\ooataufullpol=2$ and representing around 90\% of all events -- bias the neural network training, preventing it from learning a good approximation of the optimal observables. Further, there are very few good (i.e. uncorrelated) particle-level input variables related to the $\gamma\gamma\rightarrow\tau^{+}\tau^{-}$ process, which allow a multi-variate analysis to outperform simple kinematic-distribution based methods. Thus, implementing ML techniques for this task will potentially be very challenging.

\subsection{Prospects of employing optimal observables in experimental analyses}

\begin{figure}[t]
\centering
\includegraphics[height=0.45\textwidth]
{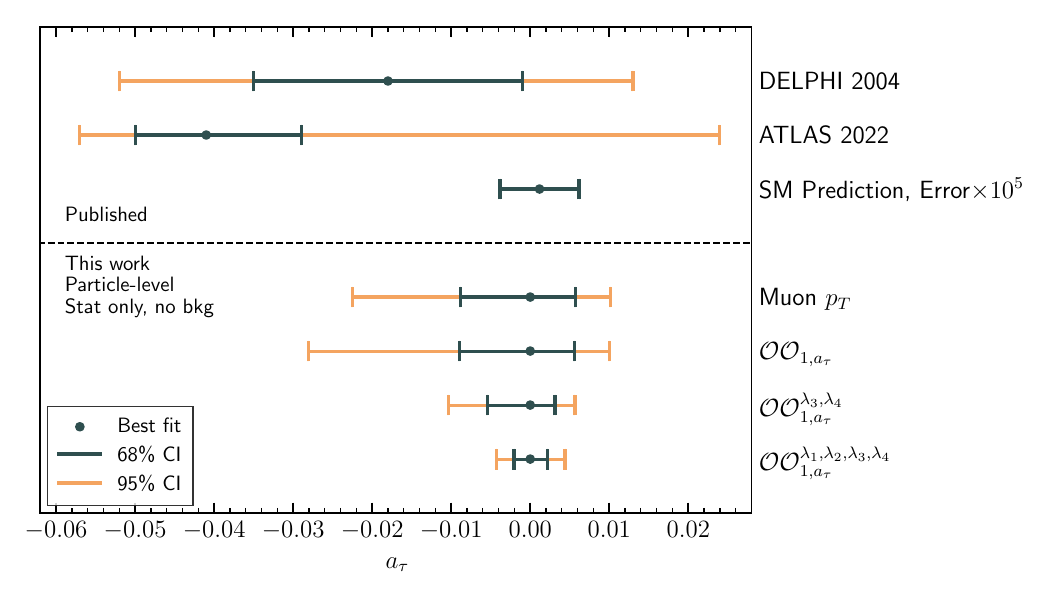}
\caption{Summary of the limits on $a_{\tau}$ obtained using optimal observables, compared the result obtained with the muon $p_T$. The published measurements are displayed for contextual reference as well and consider background contributions, full systematic uncertainties and detector-related inefficiencies, while our analysis is performed at particle-level, with no background and considering only statistical uncertainties.}\label{Fig:summary}
\end{figure}

We display the confidence intervals for $a_{\tau}$ obtained using optimal observables, together with those obtained from the muon $p_T$  employed in the existing ATLAS measurement, as well as the experimental measurement results, in Fig.~\ref{Fig:summary}. Note that our analysis is performed background-free, with statistical uncertainties only, and at particle-level, while the experimental results are determined considering background contributions, full systematic uncertainties as well as detector-related inefficiencies. Thus, the experimental results should only be used as contextual references, and should not be compared directly to the results of our analysis.

As visible in Fig.~\ref{Fig:summary}, the confidence intervals for $a_{\tau}$ can be improved with optimal observables compared to those derived using the muon $p_{T}$ observable, if knowledge of the helicities of the particles involved in the hard scattering is obtained. The fully polarized optimal observable, \ooataufullpol, requires knowledge of the helicities of the incoming photons which is inaccessible  experimentally, so its performance is, unfortunately, hypothetical only. The photon helicity averaged optimal observable, \ooatauphavg, however, only requires knowledge of the helicities of the outgoing $\tau$ leptons. The experimental challenge lies in the determination of the helicity information of the $\tau$ leptons, which is passed on to their decay products~\cite{Bullock:1992yt}, from their decay products. While there are (optimal) observables for the determination of the polarization of $\tau$ leptons 
in a sample of events\footnote{Polarization in this context relates to the asymmetry between positive and negative helicity $\tau$ leptons~\cite{Davier:1992nw}.}, there are no observables that provide exact information about the helicities of individual $\tau$ leptons on an event-by-event basis, which is required to compute the optimal observables defined in our analysis. 

Depending on the decay channel, it is possible to use energy ratio observables to determine the helicities of $\tau$ leptons on an event-by-event basis with some accuracy~\cite{Bullock:1992yt}. In $\tau^{\pm}\rightarrow\ell^{\pm}\nu_{\ell}\nu_{\tau}$ decays (where $\ell=e,\mu$), the ratio $E_{\ell^{\pm}}/E_{\tau^{\pm}}$ can be used; in $\tau^{\pm}\rightarrow\pi^{\pm}\nu_{\tau}$ decays, the best ratio is $E_{\pi^{\pm}}/E_{\tau^{\pm}}$; and in $\tau^{\pm}\rightarrow\pi^{\pm}\pi^{0}\nu_{\tau}$ decays, the best ratio is $(E_{\pi^{\pm}}-E_{\pi^{0}})/(E_{\pi^{\pm}}+E_{\pi^{0}})$. Among these, the ratio used for the $\tau^{\pm}\rightarrow\pi^{\pm}\pi^{0}\nu_{\tau}$ decay is known to have the best separation between positive and negative helicity $\tau$ leptons.

To estimate the possibility of using these variables for computing helicity-dependent optimal observables for $a_{\tau}$ in an experimental analysis, we perform a simple study in the $\left(\mu^{\pm}\nu_{\mu}\nu_{\tau}\right)\left(\pi^{\mp}\pi^{0}\nu_{\tau}\right)$ sub-region of the $1\mu 1T$ signal region. The $\tau^{\pm}\rightarrow\pi^{\pm}\pi^{0}\nu$ decay is distinguished by counting the number of neutral pions, taken directly from the simulation event record, which pass a $p_{T}>0.3$ GeV selection. The $\tau$ leptons are assigned their helicity based on a cut\footnote{A $\tau^{+}$ is assigned helicity $+$ if $E_{\mu}/E_{\tau}>0.6$ or $E_{\pi}/E_{\tau}<0.7$ depending on the decay channel, and helicity $-$ otherwise. A $\tau^{-}$ is assigned helicity $+$ if $E_{\mu}/E_{\tau}\leq0.6$ or $E_{\pi}/E_{\tau}\geq0.7$, depending on the decay channel, and helicity $-$ otherwise. The charge of the $\tau$ lepton is determined from the decay products, i.e. the $\mu^{\pm}$ and $\pi^{\pm}$.} on the energy ratios described above. The figure of merit is the joint accuracy of correctly determining the helicity of both $\tau$ leptons. We find an accuracy of 32\% for  the assignments based on the energy ratios. This is very close to a 25\% accuracy associated with random helicity assignment, given a potential additional smearing from particle reconstruction and identification inefficiencies of a real detector. The accuracy of di-$\tau$ helicity determination will be comparatively better in events where both $\tau$ leptons decay hadronically, but this was not studied here since we only consider the signal regions studied in Ref.~\cite{ATLAS:2022ryk}. Although it is technically possible, experimental realizations of the photon helicity averaged optimal observable, \ooatauphavg, will thus most likely be extremely difficult.

\section{Conclusion}\label{Sec:Conclusion}

Precise measurements of the electromagnetic dipole moments of charged leptons are powerful probes of physics beyond the Standard Model (SM) of particle physics. Recent measurements of the anomalous magnetic dipole moment (AMDM) and electric dipole moment (EDM) of the $\tau$-lepton in PbPb and proton-proton collisions by the ATLAS and CMS experiments are pushing the sensitivity closer to the SM predictions. It is thus of great interest to study avenues for optimizing the measurement strategies.

In this work, we studied the prospects of employing optimal observables for measuring the AMDM of the $\tau$-lepton, in order to achieve the strictest constraints possible. 
Optimal observables defined as matrix element ratios for new vs. SM contributions are compared to the muon $p_{T}$ observable used in the recent ATLAS measurement in PbPb collisions. The statistical analysis is based on a binned maximum likelihood fit, considering statistical uncertainties and signal only, assuming an integrated luminosity of 2.0 nb$^{-1}$, which is similar to that of the second running period of the LHC (Run 2) for PbPb collisions.

Improved measurements of the $\tau$ AMDM can be reached with a helicity-averaged optimal observable only, if a looser phase space is used than the one employed in the ATLAS measurement, or at an integrated luminosity larger than the one of Run 2. A photon helicity averaged optimal observable provides a higher sensitivity to the $\tau$ AMDM, but is experimentally very challenging due to low accuracies in determining the $\tau$ helicities from the $\tau$ decay products. Our study revealed interesting features of the matrix elements for $\gamma\gamma\rightarrow\tau^{+}\tau^{-}$ production, as well as chances and limitations of optimal observable applications in the $\gamma\gamma\rightarrow\tau^{+}\tau^{-}$ process which will be important for future experimental measurements.

The code used in this work for calculating matrix elements has been made available in a public repository~\cite{repository}. A sample script containing examples on how to use the code for computing optimal observables and reweighting events has also been provided.

\acknowledgments

We acknowledge the support by the Freiburg computing centre BFG/NEMO for the production of our event simulation samples. We thank the Baden-Württemberg Stiftung for the financial support of this research project through the Eliteprogramme for Postdocs. We are grateful for the invaluable feedback on our studies from colleagues at the University of Freiburg, in particular Markus Schumacher and Mathieu Pellen. Any errors in this publication fall under the responsibility of the authors.

\appendix

% Bibliography
\bibliographystyle{JHEP}
\bibliography{biblio.bib}
\end{document}